


%
%


\def\famname{
 \textfont0=\textrm \scriptfont0=\scriptrm
 \scriptscriptfont0=\sscriptrm
 \textfont1=\textmi \scriptfont1=\scriptmi
 \scriptscriptfont1=\sscriptmi
 \textfont2=\textsy \scriptfont2=\scriptsy \scriptscriptfont2=\sscriptsy
 \textfont3=\textex \scriptfont3=\textex \scriptscriptfont3=\textex
 \textfont4=\textbf \scriptfont4=\scriptbf \scriptscriptfont4=\sscriptbf
 \skewchar\textmi='177 \skewchar\scriptmi='177
 \skewchar\sscriptmi='177
 \skewchar\textsy='60 \skewchar\scriptsy='60
 \skewchar\sscriptsy='60
 \def\rm{\fam0 \textrm} \def\bf{\fam4 \textbf}}
\def\sca#1{scaled\magstep#1} \def\scah{scaled\magstephalf} 
\def\twelvepoint{
 \font\textrm=cmr12 \font\scriptrm=cmr8 \font\sscriptrm=cmr6
 \font\textmi=cmmi12 \font\scriptmi=cmmi8 \font\sscriptmi=cmmi6 
 \font\textsy=cmsy10 \sca1 \font\scriptsy=cmsy8
 \font\sscriptsy=cmsy6
 \font\textex=cmex10 \sca1
 \font\textbf=cmbx12 \font\scriptbf=cmbx8 \font\sscriptbf=cmbx6
 \font\it=cmti12
 \font\sectfont=cmbx12 \sca1
 \font\sectmath=cmmib10 \sca2
 \font\sectsymb=cmbsy10 \sca2
 \font\refrm=cmr10 \scah \font\refit=cmti10 \scah
 \font\refbf=cmbx10 \scah
 \def\twelverm{\textrm} \def\twelveit{\it} \def\twelvebf{\textbf}
 \famname \textrm 
\voffset=.2in \hoffset=.3in
 \normalbaselineskip=17.5pt plus 1pt \baselineskip=\normalbaselineskip
 \parindent=21pt
 \setbox\strutbox=\hbox{\vrule height10.5pt depth4pt width0pt}}


\catcode`@=11

{\catcode`\'=\active \def'{{}^\bgroup\prim@s}}

\def\screwcount{\alloc@0\count\countdef\insc@unt}   
\def\screwdimen{\alloc@1\dimen\dimendef\insc@unt} 
\def\screwbox{\alloc@4\box\chardef\insc@unt}

\catcode`@=12


\overfullrule=0pt			
\vsize=9in \hsize=6in
\lineskip=0pt				
\abovedisplayskip=1.2em plus.3em minus.9em 
\belowdisplayskip=1.2em plus.3em minus.9em	
\abovedisplayshortskip=0em plus.3em	
\belowdisplayshortskip=.7em plus.3em minus.4em	
\parindent=21pt
\setbox\strutbox=\hbox{\vrule height10.5pt depth4pt width0pt}
\def\makefootline{\baselineskip=30pt \line{\the\footline}}
\footline={\ifnum\count0=1 \hfil \else\hss\twelverm\folio\hss \fi}
\pageno=1


\def\boxit#1{\leavevmode\thinspace\hbox{\vrule\vtop{\vbox{\hrule%
	\vskip3pt\kern1pt\hbox{\vphantom{\bf/}\thinspace\thinspace%
	{\bf#1}\thinspace\thinspace}}\kern1pt\vskip3pt\hrule}\vrule}%
	\thinspace}
\def\Boxit#1{\noindent\vbox{\hrule\hbox{\vrule\kern3pt\vbox{
	\advance\hsize-7pt\vskip-\parskip\kern3pt\bf#1
	\hbox{\vrule height0pt depth\dp\strutbox width0pt}
	\kern3pt}\kern3pt\vrule}\hrule}}


\def\put(#1,#2)#3{\screwdimen\unit  \unit=1in
	\vbox to0pt{\kern-#2\unit\hbox{\kern#1\unit
	\vbox{#3}}\vss}\nointerlineskip}


\def\\{\hfil\break}
\def\newpage{\vfill\eject}
\def\center{\leftskip=0pt plus 1fill \rightskip=\leftskip \parindent=0pt
 \def\textindent##1{\par\hangindent21pt\footrm\noindent\hskip21pt
 \llap{##1\enspace}\ignorespaces}\par}
\def\unnarrower{\leftskip=0pt \rightskip=\leftskip}

%
%
%
%
%
%
%

\def\thetitle#1#2#3#4#5{
 \font\titlefont=cmbx12 \sca2 \font\footrm=cmr10 \font\footit=cmti10
  \twelverm
	{\hbox to\hsize{#4 \hfill YITP-SB-#3}}\par
	\vskip.8in minus.1in {\center\baselineskip=1.44\normalbaselineskip
 {\titlefont #1}\par}{\center\baselineskip=\normalbaselineskip
 \vskip.5in minus.2in #2
	\vskip1.4in minus1.2in {\twelvebf ABSTRACT}\par}
 \vskip.1in\par
 \narrower\par#5\par\unnarrower\vskip3.5in minus2.3in\eject}
\def\paper\par#1\par#2\par#3\par#4\par#5\par{\twelvepoint
	\thetitle{#1}{#2}{#3}{#4}{#5}} 
\def\author#1#2{#1 \vskip.1in {\twelveit #2}\vskip.1in}
\def\YITP{C. N. Yang Institute for Theoretical Physics\\
	State University of New York, Stony Brook, NY 11794-3840}
\def\WS{W. Siegel\footnote{${}^1$}{       
	\pdflink{mailto:siegel@insti.physics.sunysb.edu}\\
	\pdfklink{http://insti.physics.sunysb.edu/\~{}siegel/plan.html}
	{http://insti.physics.sunysb.edu/\noexpand~siegel/plan.html}}}


\def\sect#1\par{\par\ifdim\lastskip<\medskipamount
	\bigskip\medskip\goodbreak\else\nobreak\fi
	\noindent{\sectfont{#1}}\par\nobreak\medskip} 
\def\itemize#1 {\item{[#1]}}	
\def\vol#1 {{\refbf#1} }		 

\def\ref#1{\setbox0=\hbox{M}$\vbox to\ht0{}^{#1}$}


\def\NP #1 {{\refit Nucl. Phys.} {\refbf B{#1}} }
\def\PL #1 {{\refit Phys. Lett.} {\refbf{#1}} }
\def\PR #1 {{\refit Phys. Rev. Lett.} {\refbf{#1}} }
\def\PRD #1 {{\refit Phys. Rev.} {\refbf D{#1}} }


\hyphenation{pre-print}
\hyphenation{quan-ti-za-tion}

%
%


\def\oonoo#1#2#3{\vbox{\ialign{##\crcr
	\hfil\hfil\hfil{$#3{#1}$}\hfil\crcr\noalign{\kern1pt\nointerlineskip}
	$#3{#2}$\crcr}}}
\def\oon#1#2{\mathchoice{\oonoo{#1}{#2}{\displaystyle}}
	{\oonoo{#1}{#2}{\textstyle}}{\oonoo{#1}{#2}{\scriptstyle}}
	{\oonoo{#1}{#2}{\scriptscriptstyle}}}
\def\dt#1{\oon{\hbox{\bf .}}{#1}}  
\def\ddt#1{\oon{\hbox{\bf .\kern-1pt.}}#1}    
\def\slap#1#2{\setbox0=\hbox{$#1{#2}$}
	#2\kern-\wd0{\hfuzz=1pt\hbox to\wd0{\hfil$#1{/}$\hfil}}}
\def\sla#1{\mathpalette\slap{#1}}                
\def\bop#1{\setbox0=\hbox{$#1M$}\mkern1.5mu
	\vbox{\hrule height0pt depth.04\ht0
	\hbox{\vrule width.04\ht0 height.9\ht0 \kern.9\ht0
	\vrule width.04\ht0}\hrule height.04\ht0}\mkern1.5mu}
\def\bo{{\mathpalette\bop{}}}                        
\def~{\widetilde} 
\mathcode`\*="702A                  
\def\in{\relax\ifmmode\mathchar"3232\else{\refit in\/}\fi} 
\def\f#1#2{{\textstyle{#1\over#2}}}	   
\def\half{{\textstyle{1\over{\raise.1ex\hbox{$\scriptstyle{2}$}}}}}

\def\Gamma{\mathchar"0100}
\def\Delta{\mathchar"0101}
\def\Theta{\mathchar"0102}
\def\Lambda{\mathchar"0103}
\def\Xi{\mathchar"0104}
\def\Pi{\mathchar"0105}
\def\Sigma{\mathchar"0106}
\def\Upsilon{\mathchar"0107}
\def\Phi{\mathchar"0108}
\def\Psi{\mathchar"0109}
\def\Omega{\mathchar"010A}

\catcode`\^^?=13				    
\catcode128=13 \def €{\"A}                 
\catcode129=13 \def {\AA}                 
\catcode130=13 \def '{\c}           	   
\catcode131=13 \def ƒ{\'E}                   
\catcode132=13 \def "{\~N}                   
\catcode133=13 \def …{\"O}                 
\catcode134=13 \def †{\"U}                  
\catcode135=13 \def ‡{\'a}                  
\catcode136=13 \def ˆ{\`a}                   
\catcode137=13 \def ‰{\^a}                 
\catcode138=13 \def Š{\"a}                 
\catcode139=13 \def ‹{\~a}                   
\catcode140=13 \def Œ{\alpha}            
\catcode141=13 \def {\chi}                
\catcode142=13 \def Ž{\'e}                   
\catcode143=13 \def {\`e}                    
\catcode144=13 \def {\^e}                  
\catcode145=13 \def '{\"e}                
\catcode146=13 \def '{\'\i}                 
\catcode147=13 \def "{\`\i}                  
\catcode148=13 \def "{\^\i}                
\catcode149=13 \def •{\"\i}                
\catcode150=13 \def –{\~n}                  
\catcode151=13 \def —{\'o}                 
\catcode152=13 \def ˜{\`o}                  
\catcode153=13 \def ™{\^o}                
\catcode154=13 \def š{\"o}                 
\catcode155=13 \def ›{\~o}                  
\catcode156=13 \def œ{\'u}                  
\catcode157=13 \def {\`u}                  
\catcode158=13 \def ž{\^u}                
\catcode159=13 \def Ÿ{\"u}                
\catcode160=13 \def  {\tau}               
\catcode161=13 \mathchardef ¡="2203     
\catcode162=13 \def ¢{\oplus}           
\catcode163=13 \def £{\relax\ifmmode\to\else\itemize\fi} 
\catcode164=13 \def ¤{\subset}	  
\catcode165=13 \def ¥{\infty}           
\catcode166=13 \def ¦{\mp}                
\catcode167=13 \def §{\sigma}           
\catcode168=13 \def ¨{\rho}               
\catcode169=13 \def ©{\gamma}         
\catcode170=13 \def ª{\leftrightarrow} 
\catcode171=13 \def «{\relax\ifmmode\acute\else\expandafter\'\fi}
\catcode172=13 \def ¬{\relax\ifmmode\expandafter\ddt\else\expandafter\"\fi}
\catcode173=13 \def ­{\equiv}            
\catcode174=13 \def ®{\approx}          
\catcode175=13 \def ¯{\Omega}          
\catcode176=13 \def °{\otimes}          
\catcode177=13 \def ±{\ne}                 
\catcode178=13 \def ²{\le}                   
\catcode179=13 \def ³{\ge}                  
\catcode180=13 \def ´{\upsilon}          
\catcode181=13 \def µ{\mu}                
\catcode182=13 \def ¶{\delta}             
\catcode183=13 \def ·{\epsilon}          
\catcode184=13 \def ¸{\Pi}                  
\catcode185=13 \def ¹{\pi}                  
\catcode186=13 \def º{\beta}               
\catcode187=13 \def »{\partial}           
\catcode188=13 \def ¼{\nobreak\ }       
\catcode189=13 \def ½{\zeta}               
\catcode190=13 \def ¾{\sim}                 
\catcode191=13 \def ¿{\omega}           
\catcode192=13 \def À{\dt}                     
\catcode193=13 \def Á{\gets}                
\catcode194=13 \def Â{\lambda}           
\catcode195=13 \def Ã{\nu}                   
\catcode196=13 \def Ä{\phi}                  
\catcode197=13 \def Å{\xi}                     
\catcode198=13 \def Æ{\psi}                  
\catcode199=13 \def Ç{\int}                    
\catcode200=13 \def È{\oint}                 
\catcode201=13 \def É{\relax\ifmmode\cdot\else\vol\fi}    
\catcode202=13 \def Ê{\relax\ifmmode\,\else\thinspace\fi}
\catcode203=13 \def Ë{\`A}                      
\catcode204=13 \def Ì{\~A}                      
\catcode205=13 \def Í{\~O}                      
\catcode206=13 \def Î{\Theta}              
\catcode207=13 \def Ï{\theta}               
\catcode208=13 \def Ð{\relax\ifmmode\bar\else\expandafter\=\fi}
\catcode209=13 \def Ñ{\overline}             
\catcode210=13 \def Ò{\langle}               
\catcode211=13 \def Ó{\relax\ifmmode\{\else\ital\fi}      
\catcode212=13 \def Ô{\rangle}               
\catcode213=13 \def Õ{\}}                        
\catcode214=13 \def Ö{\sla}                      
\catcode215=13 \def ×{\relax\ifmmode\check\else\expandafter\v\fi}
\catcode216=13 \def Ø{\"y}                     
\catcode217=13 \def Ù{\"Y}  		    
\catcode218=13 \def Ú{\Leftarrow}       
\catcode219=13 \def Û{\Leftrightarrow}       
\catcode220=13 \def Ü{\relax\ifmmode\Rightarrow\else\sect\fi}
\catcode221=13 \def Ý{\sum}                  
\catcode222=13 \def Þ{\prod}                 
\catcode223=13 \def ß{\widehat}              
\catcode224=13 \def à{\pm}                     
\catcode225=13 \def á{\nabla}                
\catcode226=13 \def â{\quad}                 
\catcode227=13 \def ã{\in}               	
\catcode228=13 \def ä{\star}      	      
\catcode229=13 \def å{\sqrt}                   
\catcode230=13 \def æ{\^E}			
\catcode231=13 \def ç{\Upsilon}              
\catcode232=13 \def è{\"E}    	   	 
\catcode233=13 \def é{\`E}               	  
\catcode234=13 \def ê{\Sigma}                
\catcode235=13 \def ë{\Delta}                 
\catcode236=13 \def ì{\Phi}                     
\catcode237=13 \def í{\`I}        		   
\catcode238=13 \def î{\iota}        	     
\catcode239=13 \def ï{\Psi}                     
\catcode240=13 \def ð{\times}                  
\catcode241=13 \def ñ{\Lambda}             
\catcode242=13 \def ò{\cdots}                
\catcode243=13 \def ó{\^U}			
\catcode244=13 \def ô{\`U}    	              
\catcode245=13 \def õ{\bo}                       
\catcode246=13 \def ö{\relax\ifmmode\hat\else\expandafter\^\fi}
\catcode247=13 \def÷{\relax\ifmmode\tilde\else\expandafter\~\fi}
\catcode248=13 \def ø{\ll}                         
\catcode249=13 \def ù{\gg}                       
\catcode250=13 \def ú{\eta}                      
\catcode251=13 \def û{\kappa}                  
\catcode252=13 \def ü{\half}     		 
\catcode253=13 \def ý{\Gamma} 		
\catcode254=13 \def þ{\Xi}   			
\catcode255=13 \def ÿ{\relax\ifmmode{}^{\dagger}{}\else\dag\fi}


\def\ital#1Õ{{\it#1\/}}	     
\def\un#1{\relax\ifmmode\underline#1\else $\underline{\hbox{#1}}$
	\relax\fi}

\def\tdt#1{\oon{\hbox{\bf .\kern-1pt.\kern-1pt.}}#1}   
\def\({\eqno(}

\def\refs{\sect{REFERENCES}\par\medskip \frenchspacing 
	\parskip=0pt \refrm \baselineskip=1.23em plus 1pt
	\def\ital##1Õ{{\refit##1\/}}}


\def\õ#1{
	\screwcount\num
	\num=1
	\screwdimen\downsy
	\downsy=-1.5ex
	\mkern-3.5mu
	õ
	\loop
	\ifnum\num<#1
	\llap{\raise\num\downsy\hbox{$õ$}}
	\advance\num by1
	\repeat}
\def\upõ#1#2{\screwcount\numup
	\numup=#1
	\advance\numup by-1
	\screwdimen\upsy
	\upsy=.75ex
	\mkern3.5mu
	\raise\numup\upsy\hbox{$#2$}}



\newcount\marknumber	\marknumber=1
\newcount\countdp \newcount\countwd \newcount\countht 

%
%
\ifx\pdfoutput\undefined
\def\rgboo#1{}
\input epsf

\def\postscript#1{\special{" #1}}		
\postscript{
	/bd {bind def} bind def
	/fsd {findfont exch scalefont def} bd
	/sms {setfont moveto show} bd
	/ms {moveto show} bd
	/pdfmark where		
	{pop} {userdict /pdfmark /cleartomark load put} ifelse
	[ /PageMode /UseOutlines		
	/DOCVIEW pdfmark}
\def\bookmark#1#2{\postscript{		
	[ /Dest /MyDest\the\marknumber /View [ /XYZ null null null ] /DEST pdfmark
	[ /Title (#2) /Count #1 /Dest /MyDest\the\marknumber /OUT pdfmark}%
	\advance\marknumber by1}
\def\pdfklink#1#2{%
	\hskip-.25em\setbox0=\hbox{#1}%
		\countdp=\dp0 \countwd=\wd0 \countht=\ht0%
		\divide\countdp by65536 \divide\countwd by65536%
			\divide\countht by65536%
		\advance\countdp by1 \advance\countwd by1%
			\advance\countht by1%
		\def\linkdp{\the\countdp} \def\linkwd{\the\countwd}%
			\def\linkht{\the\countht}%
	\postscript{
		[ /Rect [ -1.5 -\linkdp.0 0\linkwd.0 0\linkht.5 ] 
		/Border [ 0 0 0 ]
		/Action << /Subtype /URI /URI (#2) >>
		/Subtype /Link
		/ANN pdfmark}{\rgb{1 0 0}{#1}}}
%
%
\else
\def\rgboo#1{\pdfliteral{#1 rg #1 RG}}

\pdfcatalog{/PageMode /UseOutlines}		
\def\bookmark#1#2{
	\pdfdest num \marknumber xyz
	\pdfoutline goto num \marknumber count #1 {#2}
	\advance\marknumber by1}
\def\pdfklink#1#2{%
	\noindent\pdfstartlink user
		{/Subtype /Link
		/Border [ 0 0 0 ]
		/A << /S /URI /URI (#2) >>}{\rgb{1 0 0}{#1}}%
	\pdfendlink}
\fi

\def\rgbo#1#2{\rgboo{#1}#2\rgboo{0 0 0}}
\def\rgb#1#2{\mark{#1}\rgbo{#1}{#2}\mark{0 0 0}}
\def\pdflink#1{\pdfklink{#1}{#1}}
\def\xxxlink#1{\pdfklink{#1}{http://arXiv.org/abs/#1}}

\catcode`@=11

\def\wlog#1{}	


\def\makeheadline{\vbox to\z@{\vskip-36.5\p@
	\line{\vbox to8.5\p@{}\the\headline%
	\ifnum\pageno=\z@\rgboo{0 0 0}\else\rgboo{\topmark}\fi%
	}\vss}\nointerlineskip}
\headline={
	\ifnum\pageno=\z@
		\hfil
	\else
		\ifnum\pageno<\z@
			\ifodd\pageno
				\tenrm\romannumeral-\pageno\hfil\lefthead\hfil
			\else
				\tenrm\hfil\righthead\hfil\romannumeral-\pageno
			\fi
		\else
			\ifodd\pageno
				\tenrm\hfil\righthead\hfil\number\pageno
			\else
				\tenrm\number\pageno\hfil\lefthead\hfil
			\fi
		\fi
	\fi}

\catcode`@=12

\def\righthead{\hfil} \def\lefthead{\hfil}
\nopagenumbers


\def\chrulefill{\rgb{1 0 0}{\hrulefill}}
\def\cdotfill{\rgb{1 0 0}{\dotfill}}
\newcount\area	\area=1
\newcount\cross	\cross=1
\def\volume#1\par{\newpage\noindent{\biggest{\rgb{1 .5 0}{#1}}}
	\par\nobreak\bigskip\medskip\area=0}
\def\chapskip{\par\ifnum\area=0\bigskip\medskip\goodbreak
	\else\newpage\fi}
\def\chapy#1{\area=1\cross=0
	\xdef\lefthead{\rgbo{1 0 .5}{#1}}\vbox{\biggerer\offinterlineskip
	\line{\chrulefill¼\hphantom{\lefthead}\chrulefill}
	\line{\chrulefill¼\lefthead\chrulefill}}\par\nobreak\medskip}
\def\chap#1\par{\chapskip\bookmark3{#1}\chapy{#1}}
\def\sectskip{\par\ifnum\cross=0\bigskip\medskip\goodbreak
	\else\newpage\fi}
\def\secty#1{\cross=1
	\xdef\righthead{\rgbo{1 0 1}{#1}}\vbox{\bigger\offinterlineskip
	\line{\cdotfill¼\hphantom{\righthead}\cdotfill}
	\line{\cdotfill¼\righthead\cdotfill}}\par\nobreak\medskip}
\def\sect#1 #2\par{\sectskip\bookmark{#1}{#2}\secty{#2}}
\def\subsectskip{\par\ifdim\lastskip<\medskipamount
	\bigskip\medskip\goodbreak\else\nobreak\fi}
\def\subsecty#1{\noindent{\sectfont{\rgbo{.5 0 1}{#1}}}\par\nobreak\medskip}
\def\subsect#1\par{\subsectskip\bookmark0{#1}\subsecty{#1}}
\long\def\x#1 #2\par{\hangindent2\parindent\rgb{0 0 1}{{\bf Exercise #1}\\{#2}}\par}
\def\refs{\bigskip\noindent{\bf \rgbo{0 .5 1}{REFERENCES}}\par\nobreak\medskip
	\frenchspacing \parskip=0pt \refrm \baselineskip=1.23em plus 1pt
	\def\ital##1Õ{{\refit##1\/}}}
\long\def\twocolumn#1#2{\hbox to\hsize{\vtop{\hsize=2.9in#1}
	\hfil\vtop{\hsize=2.9in #2}}}


\twelvepoint
\font\bigger=cmbx12 \sca2
\font\biggerer=cmb10 \sca5
\font\biggest=cmssdc10 scaled 3583

 \sca3


\def Ü{\relax\ifmmode\Rightarrow\else\expandafter\subsect\fi}
\def Û{\relax\ifmmode\Leftrightarrow\else\expandafter\sect\fi}
\def Ú{\relax\ifmmode\Leftarrow\else\expandafter\chap\fi}

\def\itemize#1 {\item{\bf#1}}
\def\itemizze#1 {\itemitem{\bf#1}}
\def\itemutem{\par\indent\indent \hangindent3\parindent \textindent}
\def\itemizzze#1 {\itemutem{\bf#1}}
\def ª{\relax\ifmmode\leftrightarrow\else\itemizze\fi}
\def Á{\relax\ifmmode\gets\else\itemizzze\fi}

\def\¢{\ominus}

\def\Ä{\varphi}  \def\¿{\varpi}

\def ò{\relax\ifmmode\cdots\else\dotfill\fi}




\def\today{\ifcase\month\or
 January\or February\or March\or April\or May\or June\or July\or
 August\or September\or October\or November\or December\fi
 \space\number\day, \number\year}

\parindent=20pt
\newskip\normalparskip	\normalparskip=.7\medskipamount
\parskip=\normalparskip	



\catcode`\|=\active \catcode`\<=\active \catcode`\>=\active 
\def|{\relax\ifmmode\delimiter"026A30C \else$\mathchar"026A$\fi}
\def<{\relax\ifmmode\mathchar"313C \else$\mathchar"313C$\fi}
\def>{\relax\ifmmode\mathchar"313E \else$\mathchar"313E$\fi}



\pageno=0

\paper

\biggest{\rgb{1 0 .5}{SUPERWAVES}}

\author\WS\YITP

02-33

June 17, 2002

We give some supersymmetric wave solutions, both chiral (selfdual) and nonchiral,
to interacting supersymmetric theories in four dimensions.

\pageno=2

Ü1. Introduction

Exact wave solutions to interacting theories [1] have recently drawn attention
for application to the AdS/CFT correspondence [2].  In Yang-Mills theory they 
take the form (neglecting matter sources)
$$ dx^m A_m = dx^+ A_+(x^+,x^i),ââ(»^i)^2 A_+ = 0 $$
and in gravity
$$ dx^m dx^n g_{mn} = dx^m dx^n ú_{mn} 
	+dx^+ dx^+ g_{++}(x^+,x^i),ââ(»^i)^2 g_{++} = 0 $$
In the absence of sources, they describe free particles in interacting theories;
in the presence of sources, they describe free particles produced by the
interaction of particles of lower spin.
In this paper we consider supersymmetric generalizations of such solutions.
They describe particles of various spins, related by supersymmetry; 
in some solutions there are no interactions, in others, some act as sources
for the highest spin.

Ü2. Complex solutions

In quantum field theory one begins with a classical solution to the field equations
as a vacuum, and perturbs about it:  The perturbations include both tree and loop graphs.
Generally real Euclidean solutions are considered, as appropriate to the Euclidean
path integral.

Two types of real Euclidean solutions to interacting field theories have been considered:
(1) Those that have time-reversal invariance with respect to the coordinate to be
Wick rotated remain real after analytic continuation to Minkowski space.  
(The symmetry $t£-t$ becomes complex conjugation invariance $it£-it$, i.e., reality.)
This includes static solutions, such as monopoles or spherical black holes.
(But see, e.g., the use of Euclidean space to eliminate black hole singularities [3].)
Nonstatic solutions include certain cosmological ones, with appropriate choice of
time coordinate.
(Euclidean space has also been proposed to eliminate singularities in cosmology [4].)

(2) More complicated time-dependent solutions, such as instantons, can become
complex upon continuation, so their validity has sometimes been questioned.
(Instantons are complex in Minkowski space because there selfduality is a complex
condition on the field strength.)
Since wave solutions by definition propagate at the speed of light (as distinguished
from solitons, etc.), we need to consider the general validity of such complex solutions.
We begin by looking at such solutions that are themselves perturbative.

Since we are considering classical solutions, such perturbative solutions correspond
to tree graphs.  It is generally stated that tree graphs of quantum field theory
are equivalent to classical field theory [5].  However, there are two important
differences:  
(1) Classical field theory requires retarded (or advanced) 
propagators to preserve reality, while quantum field theory requires the complex
St¬uckelberg-Feynman propagator.  (In time-3-momentum space,
compare the retarded $2Ï(t)sin(¿t)$ or advanced $-2Ï(-t)sin(¿t)$ to the
St¬uckelberg-Feynman
$ie^{-i¿|t|}$.)  This requirement is equivalent to analytic continuation
from real solutions in Euclidean space.  (Consider $ie^{-i¿|t|}$ vs.¼$e^{-¿|t|}$
under complex conjugation, including $\vec p £ -\vec p$.)

(2) While classical field theory uses the same real field for each external line,
quantum field theory uses different one-particle wave functions for each line, and 
they are complex in Minkowski space, from choosing either positive (initial) 
or negative (final) energies for each line, by the same considerations
as for the propagator.  (Similar remarks apply for theories with multiple fields, or with
respect to reality properties for complex fields, which can always be re-expressed in
terms of real ones.)  But any such solutions that are, e.g., traveling in a given direction
are also complex in Euclidean space, since spatial momentum is not Wick rotated.
(However, such solutions can be real in two space and two time dimensions.)
Thus, external line factors can result in complex solutions in both Minkowski 
and Euclidean space in quantum field theory.  In particular, for solutions
propagating at the speed of light it is not possible to eliminate the problem by
choice of time axis.  (In classical field theory, adding
the complex conjugate solution to make a real solution can still give a wave
propagating in a given direction, since changing the signs of both the momentum
and energy leaves the velocity invariant.)

Thus we are led to consider only complex solutions that are consistent
with the requirements of perturbation theory.  
In fact, wave solutions are perturbative:  Their
contributions to the field are linear in the coupling, and satisfy the 
non-self-interacting Laplace equation (perhaps with sources)
in two fewer dimensions; they take the exact same form
as in the Abelian theory.  (Yang-Mills wave solutions are the same as Abelian ones;
gravitational wave solutions are the same as the flat metric plus an Abelian
spin-2 field.)  However, this requirement generalizes to
nonperturbative solutions, by analytic continuation from regions of
spacetime where they can be expressed perturbatively.  (For example,
the Schwarzschild solution can be derived perturbatively, at least outside
the event horizon [6].)

Ü3. Selfdual solutions

In particular, in four dimensions these wave solutions to the two-dimensional
Laplace equation decompose into the sum of parts analytic and antianalytic
in the complex coordinates for those two dimensions.  Treated as a classical
theory, the analytic and antianalytic terms must be complex conjugates to
preserve reality; but in quantum field theory the two terms correspond to
positive and negative helicity, and can be treated as independent.
In practice, we set one term to vanish to describe a particle of definite
helicity.

Truncation to states of given helicity corresponds directly to truncation of
the theory to the selfdual theory, as has been demonstrated both in
Feynman diagrams [7] and directly in the action [8].  
(This differs from early actions for selfdual theories [9], which not only did not
correspond to truncations of nonselfdual theories, but violated Lorentz
invariance and even dimensional analysis.  Also, when coupling selfdual
gravity to selfdual Yang-Mills, they violated Lorentz invariance and selfduality
of the field equations themselves.)  Here we will concentrate
our attention to maximally symmetric theories, since in the lightcone
they can be described by a single (super)field.

In the lightcone, the action for (super) Yang-Mills theory can be written as
the sum of (1) the kinetic term, (2) the selfdual (3-point) vertex, (3) the
antiselfdual vertex, and (4) the 4-point vertex [10].   The (polynomial form of)
the lightcone gauge for the selfdual theory [11] is described in terms of the
same fields, but keeping just the first two terms in the action [12]:
$$ S = trÇd^4 x¼d^4 ϼ[üÄõÄ +g\f13 Ä(»_{[-}Ä)(»_{t]}Ä)] $$
where $»_t$ is the derivative with respect to the complex coordinate.
Since any solution to the selfdual theory is also a solution to the full theory,
it is sufficient to consider this truncated action.
Since the action contains no derivatives with respect to the anticommuting
coordinates, the field equations take the same form for all spins
(``spectral flow"):  Only the labels on the component fields change,
since the helicities at the vertex must add to 1.  Consequently, the wave
solution is ÓidenticalÕ to that of nonsupersymmetric selfudal
Yang-Mills theory (i.e., one helicity of the nonselfdual theory), except
that the field has arbitrary dependence on the anticommuting coordinates.
Since the field strength is defined by taking the same derivatives
($»_-$ and $»_t$) that appear in the action, not both of them can vanish.
We will choose $»_-$ to vanish; the other choice is related by parity.
Furthermore, since $üõ=-»_+ »_- +»_t »_{Ðt}$, the wave solution is then
$$ Ä(x,Ï) = Ä(x^+,x^t,Ï^î) $$
 and the only nonvanishing components of the gauge field and field strength are
$$ A_+ = »_t Ä,ââF_{t+} = »_t^2 Ä $$
Except for the arbitrary $Ï$ dependence, this is the same as the wave
solution stated in the introduction, specialized to one of the two chiralities
in D=4 (analytic, and not antianalytic).  These superfields yield the
corresponding component fields at $Ï=0$; other component fields
(for spinors and scalars) reside in $Ä$ at higher orders in $Ï$
($Æ_î=»_î »_t Ä$, $\Ä_{îû}=»_î »_û Ä$, etc.).

Similar remarks apply to (super)gravity, except that the action [13] is 
polynomial only in the lightcone gauge [14]:
$$ S = Çd^4 x¼d^8 ϼ[üÄõÄ +û\f16 i Ä(»_{[-}»_{[-}Ä)(»_{t]}»_{t]}Ä) 
	+g\f16 Ä(»_î »_{[-}Ä)(»_{t]}»_î Ä)] $$
where the last term is an optional gauge coupling for gauged
selfdual supergravity, for gauge group SO(8) (but contractions and
Wick rotations of this group are also allowed).  The wave solutions
$$ Ä(x,Ï) = Ä(x^+,x^t,Ï^î);ââg_{++} = »_t^2 Ä,ââR_{t+t+} = »_t^4 Ä $$
exhibit the usual ``direct product" structure in terms of the (super)
Yang-Mills solutions.

These results can be generalized to nonmaximal supersymmetry:
There the action takes the same form, but one field in each term is
replaced with a second, Lagrange multiplier field (since it then
appears linearly).  The equations of motion (and thus the solutions)
for the original (selfdual) field are identical, while those for the
Lagrange multiplier field can be solved by setting it to zero.
(For the number of supersymmetries $N$ one less than maximal,
which describes a theory equivalent to the maximal, the counting
of nonvanishing components is the same because of the
existence of a reality condition on the field only for the maximal case.)

Ü4. Nonselfdual solutions

Although complex solutions are allowed, and may be simpler, real
solutions are not prohibited:  In Minkowski space, they correspond 
to identical incoming and outgoing states, and thus have a simpler
classical interpretation.  In the supersymmetric case such wave
solutions are not free; such solutions have already been considered
for nonsupersymmetric waves (as bosonic solutions in supersymmetric
theories; see, e.g., [2]).

Here we consider the simplest case in D=4, namely N=1 super
Yang-Mills.  We first consider the theory in components:
In two-component spinor notation, the nonvanishing component
of the vector is $A_{+À+}$.  Therefore, we choose for the nonvanishing
spinor components $Æ_+$ and its complex conjugate $ÐÆ_{À+}$.
Again dropping dependence on $x^{-À-}$, the field equations reduce to
$$ »_{-À+}Æ_+ = »_{+À-}ÐÆ_{À+} = 0,ââ»_{+À-}»_{-À+}A_{+À+} = iÐÆ_{À+}Æ_+ $$
We can then write the solution in terms of
$$ ÐÄ_{À+}(x^{+À+},x^{+À-}),âÄ_{+À+}(x^{+À+},x^{+À-}) $$
(where $x^{+À-}$ is $x^t$) as
$$ Æ_+ = »_{+À-}ÐÄ_{À+},âÐÆ_{À+} = »_{-À+}Ä_+;ââ
	A_{+À+} = Ä_{+À+} +ÐÄ_{+À+} +iÄ_+ ÐÄ_{À+} $$

The superfield formulation is interesting in its own right.
It is not as simple as in the selfdual case, but may suggest
generalizations.  As in the component analysis, the only
nonvanishing components of the spinor field strength are
$W_+$ and $ÑW_{À+}$.  In the chiral representation, the only
nontrivial covariant derivatives are then $á_+$ and $á_{+À+}$.
We then have
$$ á_+ = e^{-V}d_+ e^V;ââd_- V = Ðd_{À-}V = »_{-À-}V = 0 $$
$$ W_+ = Ðd_{À+}e^{-V}»_{+À-}e^V $$
The field equations are then
$$ á^Œ W_Œ ¾ »_{-À+}e^{-V}»_{+À-}e^V = 0 $$
These are familiar as the field equations for the (Euclidean)
2D Wess-Zumino model, with solution
$$ e^V = e^{Я(x^{+À+},x^{-À+})}e^{¯(x^{+À+},x^{+À-})} $$
 where $¯$ (as $V$) effectively depends on just $Ï^+$ and $ÐÏ^{À+}$
out of the anticommuting coordinates.  Component expansion
then reproduces the above results (where we set the $Ï=ÐÏ=0$
component of $¯$ and thus $V$ to vanish in a Wess-Zumino gauge).

Ü5. Future directions

An interesting application of perturbation about selfdual wave solutions
would be a description of the full theory by a perturbation in helicity.
This could be particularly useful for gravity, where only the selfdual
action is polynomial.  

Generalizations to higher dimensions might be
relevant for further generalizations of the AdS/CFT correspondence.
For example, for super Yang-Mills in D=3,4,6,10, where there are no
scalars, we again keep only the $A_+$ component of the vector, and
just the $©_+ Æ$ part of the spinor, to find
$$ ©_i »_i Æ = 0,ââ( »_i)^2 A_+ = ÐÆ©_+ Æ $$
 so that a free solution to the Dirac equation in the transverse
dimensions acts as source to the Abelian $A_+$, as we have already seen
for D=4 (and is somewhat trivial in D=3, where such wave solutions reduce
to the usual plane waves).

ÜAcknowledgments

I thank Martin Ro×cek for pointing out references [3-4].
This work was supported in part by the National Science Foundation
Grant No.¼PHY-0098527.

\refs

£1 H.W. Brinkmann, ÓMath. AnnalenÕ É94 (1925) 119;\\
	I. Robinson, unpublished lectures (1956);\\
	J. H«ely, ÓCompt. Rend. Acad. Sci.Õ É249 (1959) 1867;\\
	A. Peres, \xxxlink{hep-th/0205040}, \PR 3 (1959) 571;\\
	J. Ehlers and W. Kundt, Exact solutions of gravitational field equations,
	in ÓGravitation: An introduction to current researchÕ, ed. L. Witten
	(Wiley, 1962) p. 49
£2 M. Blau, J. Figueroa-O'Farrill, C. Hull, and G. Papadopoulos,
	\xxxlink{hep-th/0110242}, ÓJHEPÕ É0201 (2002) 047,
	\xxxlink{hep-th/0201081}, ÓClass. Quant. Grav.Õ É19 (2002) L87;\\
	R.R. Metsaev, \xxxlink{hep-th/0112044}, \NP 625 (2002) 70;\\
	D. Berenstein, J.M. Maldacena, and H. Nastase, \xxxlink{hep-th/0202021},
	ÓJHEPÕ É0204 (2002) 013
£3 J.B. Hartle and S.W. Hawking, \PRD 13 (1976) 2188
£4  J.B. Hartle and S.W. Hawking, \PRD 28 (1983) 2960
£5 Y. Nambu, \PL 26B (1968) 626;\\
	D.G. Boulware and L.S. Brown, ÓPhys. Rev.Õ É172 (1968) 1628
£6 M.J. Duff, \PRD 7 (1973) 2317
£7 W.A. Bardeen, ÓProg. Theor. Phys. Suppl.Õ É123 (1996) 1;\\
	D. Cangemi, \xxxlink{hep-th/9605208}, \NP 484 (1997) 521
£8 G. Chalmers and W. Siegel, \xxxlink{hep-th/9606061}, \PRD 54 (1996) 7628
£9 H. Ooguri and C. Vafa, ÓMod. Phys. Lett.Õ A5 (1990) 1389, \NP 361 (1991)
	469, É367 (1991) 83;\\
	N. Marcus, \xxxlink{hep-th/9207024}, \NP 387 (1992) 263
£10 S. Mandelstam, \NP 213 (1983) 149;\\
	L. Brink, O. Lindgren, and B.E.W. Nilsson, \NP 212 (1983) 401
£11 A.N. Leznov, ÓTheor. Math. Phys.Õ É73 (1988) 1233,\\
	A.N. Leznov and M.A. Mukhtarov, ÓJ. Math. Phys.Õ É28 (1987) 2574;\\
	A. Parkes, \xxxlink{hep-th/9203074}, \PL 286B (1992) 265
£12 W. Siegel, \xxxlink{hep-th/9205075}, \PRD 46 (1992) R3235
£13 W. Siegel, \xxxlink{hep-th/9207043}, \PRD 47 (1993) 2504
£14 J.F. Pleba«nski, ÓJ. Math. Phys.Õ É16 (1975) 2395

\bye